# Building hierarchical martensite

Stefan Schwabe[1,2], Robert Niemann[1], Anja Backen[1], Daniel Wolf[1], Christine Damm[1], Tina Walter[1], Hanus Seiner[3], Oleg Heczko[4], Kornelius Nielsch[1,2], and Sebastian Fähler[1,*]

[1] Leibniz IFW Dresden, Helmholtzstraße 20, 01069 Dresden, Germany.
[2] TU Dresden, Institute of Materials Science, 01062 Dresden, Germany.
[3] Czech Academy of Sciences, Institute of Thermomechanics, 18200 Prague, Czech Republic.
[4] Czech Academy of Sciences, Institute of Physics, 18221 Prague, Czech Republic.
*e-mail: s.faehler@ifw-dresden.de

**Martensitic materials show a complex, hierarchical microstructure containing structural domains separated by various types of twin boundaries. Several concepts exist to describe this microstructure on each length scale, however, there is no comprehensive approach bridging the whole range from the nano- up to the macroscopic scale. Here, we describe for a Ni-Mn-based Heusler alloy how this hierarchical microstructure is built from scratch with just one key parameter: the tetragonal distortion of the basic building block at the atomic level. Based on this initial block, we introduce five successive levels of nested building blocks. At each level, a larger building block is formed by twinning the preceding one to minimise the relevant energy contributions locally. This naturally explains the occurrence of different types of twin boundaries. We compare this scale-bridging approach of nested building blocks with experiments in real and reciprocal space. Our approach of nested building blocks is versatile as it can be applied to the broad class of functional materials exhibiting diffusionless transformations.**

Twin boundaries (TBs) connecting different orientations of the unit cell are the characteristic feature of a martensitic microstructure. This microstructure forms after a diffusionless structural transformation from a high temperature austenite to a low temperature martensite phase. Commonly, TBs are observed at all length scales, from the nano- up to the macroscale. Such a hierarchical, twins-within-twins microstructure is found in many different materials including high strength martensitic steel[1] or NiTi as a prototype shape memory alloy[2,3]. A similar microstructure can also be observed in several ferroelectric[4] and multiferroic[5] materials. Recently, martensitic systems have also emerged as energy materials with Ni-Mn-based Heusler alloys[6,7] being of particular interest. They can be used to convert waste heat into electricity by thermomagnetic energy harvesting[8,9] or provide a more energy efficient cooling using magnetocaloric[10], elastocaloric[11] or multicaloric effects[12]. Furthermore, they are utilised for high stroke actuation by either magnetically induced reorientation[13,14] or a magnetically induced phase transformation[15].

Despite the broad range of hierarchical materials, no comprehensive approach exists that can describe the crystallographic features of the martensitic microstructure across all length scales. Most descriptions – and

experiments – consider just one length scale, e.g. the adaptive concept[16] describes merely one part of the nanoscale, and models based on nonlinear elasticity[17] describe only the µm-scale and above. Similar limitations also hold for other methods like density functional theory[18] or molecular dynamics[19]. Although for particular length scales a good agreement between theory and experiment exists[20-22], there is only one approach explaining all length scales in a hierarchical microstructure[23]. It was suggested that a stepwise compensation of each strain component occurs at every level. However, recent in situ experiments indicate that this concept of energy equilibrium is not appropriate[7], which leads to the question: Is a scale-bridging description of the experimentally observed hierarchical martensitic microstructure possible? It should be simple, originating from some fundamental properties of the lattice.

To resolve this question, we chose the well-studied Heusler alloy Ni-Mn-Ga as our model system and investigated it comprehensively by looking at all length scales within the real space (section: The quest) as well as reciprocal space (section: Bonus level). This enables us to answer two additional questions: Which energy minimisation processes lead to the occurrence of all these different types of TBs, and can "building blocks" be identified that are connected by these TBs at each length scale?

Starting with the basic tetragonal unit cell that results from the cubic to tetragonal structural transition, our approach considers five levels built upon one another. It allows a seamless connection of nano-twinning proposed in the adaptive concept[16] with continuum mechanics[17,24] by using the recent concept of ordering nanotwins[25]. In Fig. 1, an overview of our model is sketched, which acts as an outline of this paper. On each level, a new building block is introduced, which is used to construct the next higher level, similar to a computer game. Every single level is explained in detail after the experimental part. With this, we can show that the distortion of the tetragonal building block on level 0 is the key parameter sufficient for defining most of the martensitic microstructure.

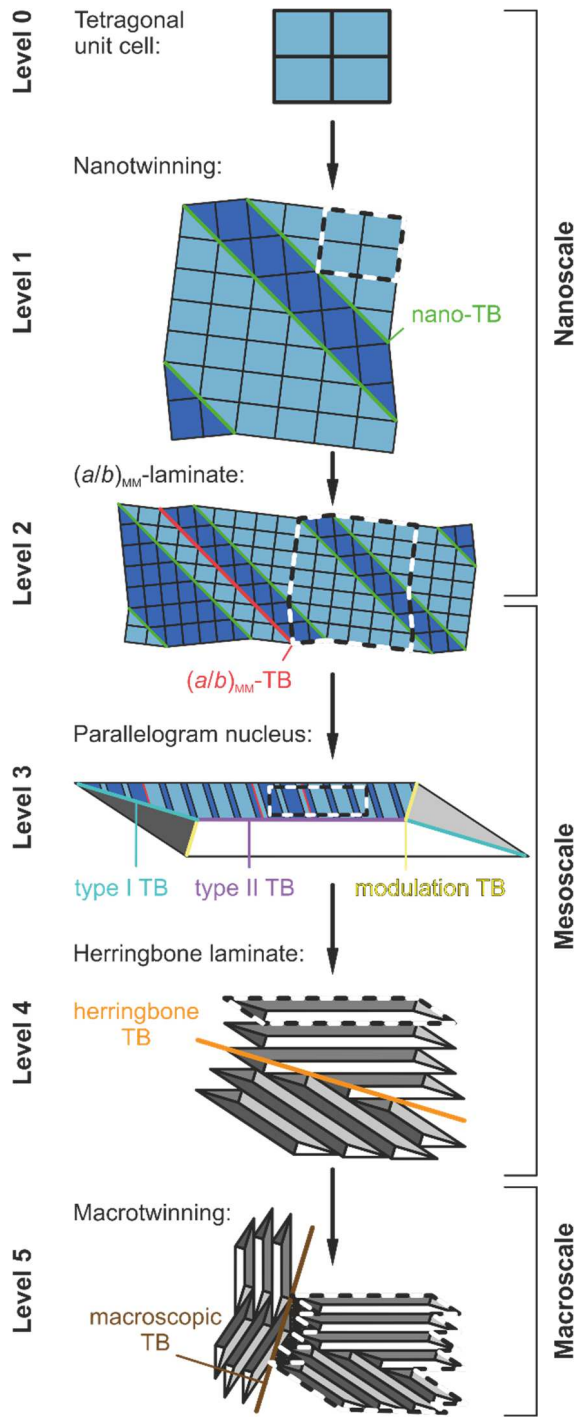

**Fig. 1 Overview of nested building blocks and twin boundaries (TB) from the nano- to the macroscale.** The microstructure development is classified into levels, starting with the basic tetragonal cell at "level 0". In each subsequent level, a new building block is introduced. It is constructed by combining the building blocks from the previous level by twinning (highlighted in different colours). The levels can be assigned to their typical length scales from the nano- to the macroscale.

**The quest: Identifying twin boundaries and building blocks on all length scales**

Before summarising the experimental findings, we introduce some general terms. A hierarchical microstructure has several common features at all length scales. For each level shown in Fig. 1, we will identify the smallest *building block* sufficient for the construction of the martensitic microstructure at that level. Regions with equally oriented building blocks are called *variants*. A *twin boundary* (TB) connects two variants of different orientation. The TB must satisfy the so-called condition of kinematic compatibility[24], which ensures that it can exist without long range stress fields. According to the symmetry operation connecting both variants, three different symmetries of TBs are established: type I TBs, type II TBs and compound TBs described in more detail in ref.[24]. Furthermore, we will use the term *laminate* for regions of parallel TBs. As these terms are used at all levels, the associated length scale will always be included.

To illustrate the hierarchical microstructure, we use an epitaxial Ni-Mn-Ga film grown on MgO (001) as a model system. Similar to single crystals, this has the advantage that its martensitic microstructure is not disturbed by grain boundaries within the austenite. In addition, the MgO substrate provides a fixed reference frame. As a starting point, we sort all the TBs into the different levels and identify the building blocks. For this purpose, a zoom-in into the hierarchical microstructure is shown in Fig. 2. According to the common approach, here we start at the macroscopic scale working our way towards the nanoscale. The scanning electron microscopy (SEM) image in Fig. 2a (see also Fig. S1 in the supplementary for a larger area) shows a macroscopic top view. It displays two notably different microstructures of almost parallel lines. We call them type X (lines under 45° to the picture borders) and type Y (lines parallel to the picture border)[26]. They are also known as high contrast zone and low contrast zone[27]. At the largest length scale, macroscopic TBs[28] (brown) are apparent (cf. also Fig. 1, level 5), which are sometimes also called colony boundaries[6]. They occur where two differently aligned type X and/or type Y zones meet. Most of the visible lines within type X and type Y have been identified as mesoscopic type II TBs (purple) with some type I TBs (cyan) also present[29]. In type X areas, a herringbone laminate is visible (level 4), incorporating herringbone TBs (orange). The characteristics of the mesoscopic TBs become clearer when looking at the transmission electron microscopy (TEM) image of the cross-section through type Y martensite (Fig. 2b) marked by a red line in Fig. 2a. Three types of TBs can be distinguished[7] differing by their angle towards the substrate: type I TBs (cyan), type II TBs (purple) and modulation TBs (yellow), which are of compound type. Martensitic nuclei are sketched in white in an intermediate growth stage, when they just arrive at the substrate, described in more detail on level 3.

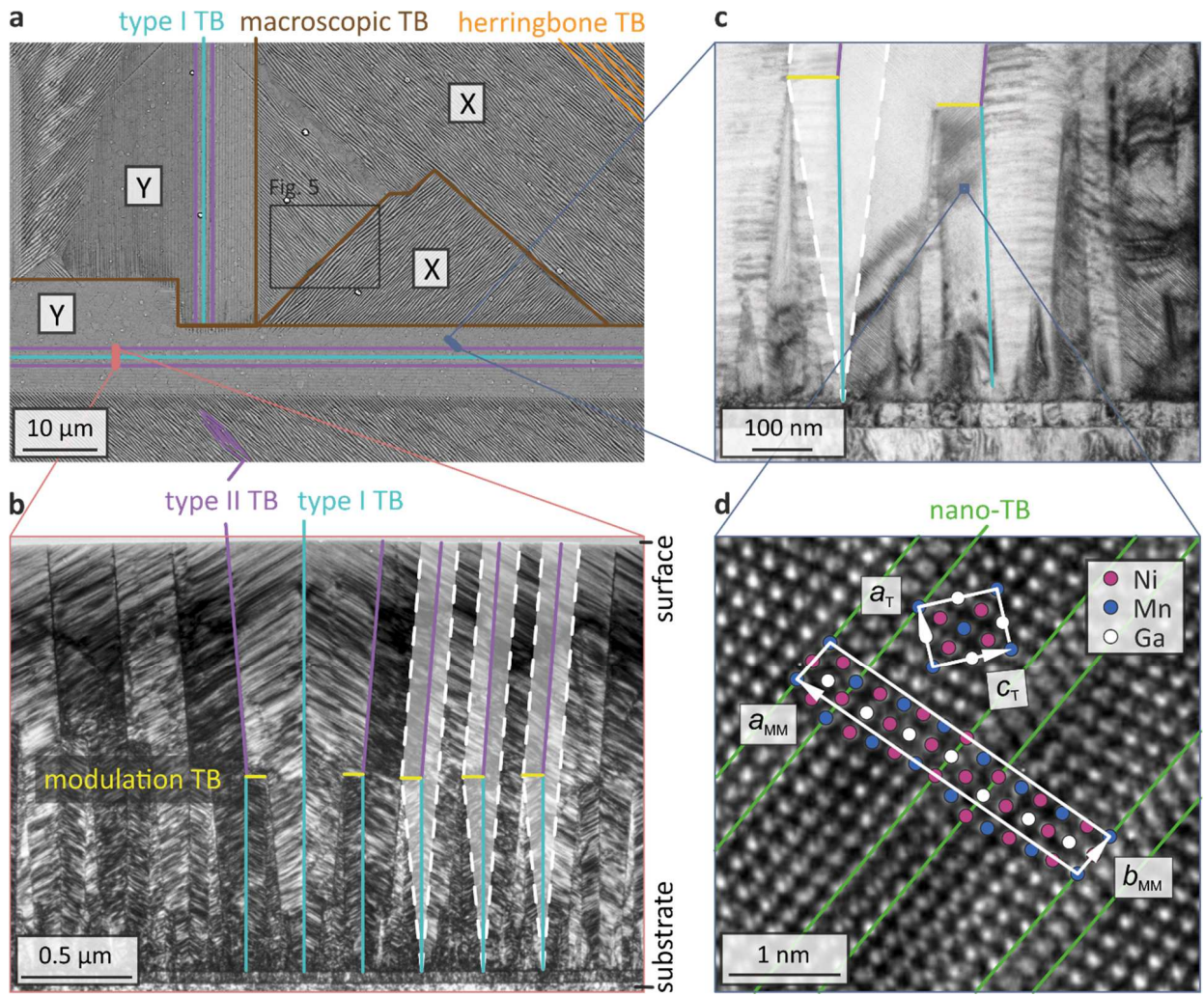

**Fig. 2 Zoom into a hierarchical microstructure with six different types of twin boundaries (TB), which are selectively highlighted. a** Macroscopic top view showing areas of differently aligned laminates (type X and Y) observed by SEM (backscattered electron contrast). These laminates are connected by macroscopic TBs (brown) and incorporate mesoscopic TBs, mainly type II (purple) and a few type I TBs (cyan). Type II TBs form a characteristic herringbone laminate discussed in more detail on level 4 observing the area framed in black. **b** The mesoscopic scale revealed by bright-field (BF)-TEM of a few tens nanometre thin cross-section sample (position marked red in **a**). The important TBs at this length scale are type I, type II and modulation TBs (yellow), which are of compound type. Martensitic nuclei are sketched in white in an intermediate growth stage, when they just reach the substrate. The dashed white lines denote the interface to the austenite at this moment (not visible anymore in the fully transformed sample). **c** BFTEM image of a second cross-section cut rotated by 45° with regard to (**b**). This enables us to see the modulations at the nano-scale. **d** HR-TEM image revealing the periodically arranged nano-TBs (green) zooming into the area marked with a blue square in (**c**). A modulated, monoclinic unit cell (with in-plane axes $a_{MM}$ and $b_{MM}$) is marked as well as the tetragonal building block (axes $a_T$ and $c_T$). The atom mapping is chosen arbitrary as all columns of atoms have almost the same contrast. The figure edges in (**a**) are parallel to $[110]_A$ and $[\bar{1}10]_A$, respectively. This figure without the overlays is available in the supplementary (Fig. S2).

A TEM image of a second cross-section (blue), cut 45° rotated in-plane compared to Fig. 2b, is shown in Fig. 2c. It allows identifying the features at the atomic scale as shown in a further high resolution (HR)-TEM zoom in (Fig. 2d). The investigated region is marked by a blue square in Fig. 2c. As reported in detail by various other groups[30,31], at this smallest length scale modulations become visible, which can be identified as nano-TBs[16] (green). Nano-TBs connect differently aligned orientations of a tetragonal building block of a non-modulated martensite with the axes $a_T$ and $c_T$ (level 1)[32]. The ratio of both axes ($c_T/a_T$-ratio) is the key parameter used for our model, as it characterizes the simplest building block. For our particular sample, five atomic planes shift in one direction followed by two planes shifting in the other direction. Preserving the chemical order, this is a 14M modulated martensite (MM) exhibiting a $(5\bar{2})_2$ modulation in Zhdanov notation (Fig. 2d).

**Level 1: Formation of a modulated unit cell by nanotwinning**

We start our description at the atomic scale with the fundamental transition between austenite (lattice parameter $a_A$) and tetragonal martensite (lattice parameters $a_T$ and $c_T$) as sketched in Fig. 3a. This transition occurs due to the lower free energy of the tetragonal phase below the transition temperature. The tetragonal distortion $c_T/a_T$ can be estimated from DFT calculations and is typically around 1.25 for Ni$_2$MnGa[33,34]. However, as these calculations commonly consider the situation at zero Kelvin, they usually overestimate the value. In addition, the $c_T/a_T$-ratio can be influenced by chemical composition and order, and thus it differs for different samples.

In the following, we use this ratio as our key parameter to construct a modulated unit cell. Generally, the tetragonal distortion can occur in each of the three spatial directions. However, on this level it is sufficient to consider only the two tetragonal cells sketched in dark and light blue colour in Fig. 3a. For a further simplification, it is convenient to subdivide the austenite and the tetragonal martensite cells into smaller "building blocks". These smaller blocks have just half the lattice parameters of chemically ordered Heusler alloys[32].

The transition is of first order, therefore austenite and martensite must coexist. They form a compatible phase boundary between them, called "habit plane", which ideally has a minimum of excess interface energy. As martensitic transitions are diffusionless, the number of building blocks on both sides of this habit plane must be equal. If a single orientation of tetragonal building blocks were connected to the same number of austenite ones, there would be a huge elastic deformation due to the large lattice misfit. To reduce this misfit and thus the elastic energy, it is favourable to combine a particular ratio of tetragonal blocks differing in the direction of their long axis (light and dark blue in Fig. 3c). This is done by introducing nano-TBs (Fig. 3b) requiring only low excess energy. The resulting twinned arrangement forms a compatible habit plane (sketched in Fig. 3c). On both sides of the habit plane, the number of building blocks as well as their total length is equal.

The distance between TBs in this nano-laminate can be determined by applying continuum theory[16]: the sum of excess energy of all TBs together with the elastic energy of a habit plane must be minimised. According to the adaptive concept[16], a low specific TB energy and a large shear modulus can favour a reduction of the distance between TBs down to the finite size of the building blocks. This is applicable for the particular Ni$_2$MnGa system[35], which explains the occurrence of nano-TBs as the first level of TBs, having a very narrow spacing.

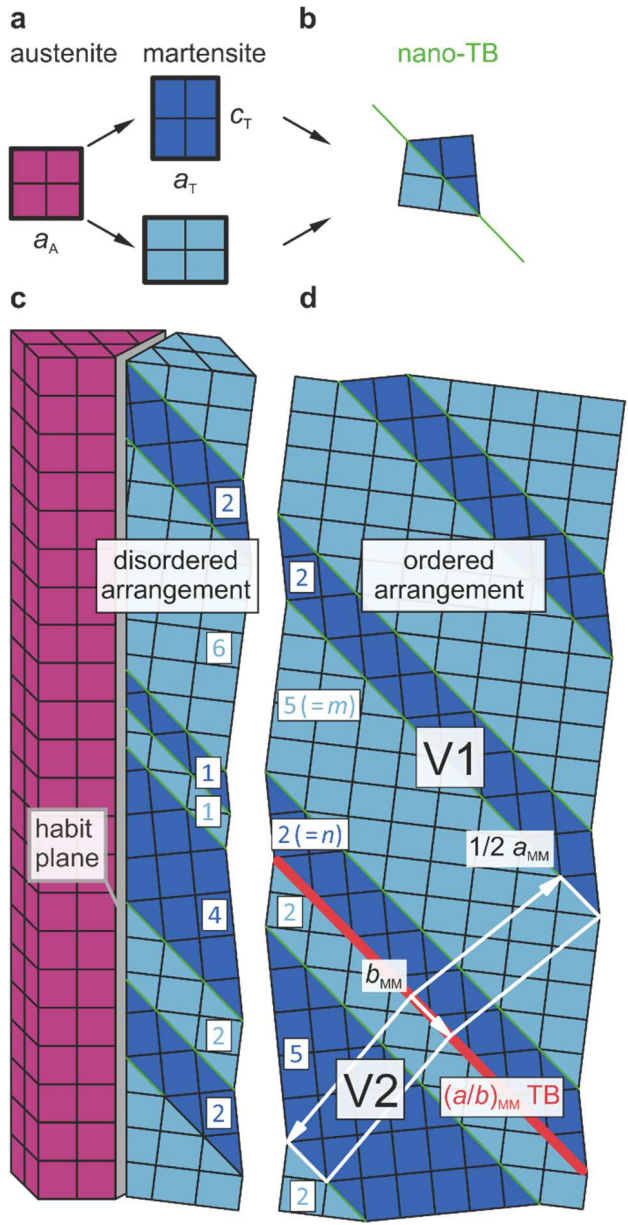

**Fig. 3 Construction of the building blocks for level 1 (modulated unit cell) and level 2 (($a/b$)$_{MM}$-laminate) using the basic tetragonal cell. a** 2D projection of the fundamental phase transition from the cubic austenite (magenta) to the tetragonal martensite (blue) when cooling below the martensitic transition temperature. The tetragonal building blocks can occur in two orientations, depicted in light and dark blue. **b** Twin boundaries (TB) between both orientations of the tetragonal unit cell are called nano-TBs (green line). **c** Disordered arrangement of nano-TBs forming a habit plane to the austenite. To illustrate that at the nano scale some disturbance of the lattice occurs, the habit plane is drawn as a grey region. **d** Sketch of the consequence of the interaction energy between nano-TBs leading to a transformation to a periodic (ordered) arrangement (($5\bar{2}$)$_2$-stacking). Because of this ordering process, a modulated structure as well as ($a/b$)$_{MM}$-TBs form (red line). The regions that are separated by ($a/b$)$_{MM}$-TBs are again variants according to our definition. The ($a/b$)$_{MM}$-TB is a mirror plane between variant 1 (V1) and V2. They have a monoclinic, modulated unit cell (surrounded by white lines). In $a_{MM}$-direction, only half of this cell is sketched. To account for the chemical order in Heusler alloys, the unit cell would be twice as long.

In the textbook case[36], an irregular arrangement of nano-TBs (cf. Fig. 3c) can minimise the elastic energy at the habit plane as they just have an excess energy. Recently however, an additional interaction energy was introduced[25] explaining the quite periodic modulations observed in experiments (cf. Fig. 3d). In our sample, a ($5\bar{2}$)$_2$-stacking is present. At level 1, we consider only the upper part of Fig. 3d (above the red line), which we will call "variant 1" (V1).

We assign $n$ to be the number of building blocks with their long axis approximately parallel to the habit plane (dark blue). The number of light blue blocks is specified by $m$. As in our martensitic Heusler alloy the $c_T/a_T$-ratio is > 1, one needs more light blue than dark blue building blocks for length conservation, meaning $m > n$. According to the concept of ordering nanotwins[25], $n = 2$ is favoured due to a minimum of interaction energy between the nano-TBs. To determine $m$, we consider that at the habit plane the number of unit cells and their length for austenite and martensite must be equal:

$$(n + m) \cdot a_A = n \cdot c_T + m \cdot a_T$$

To solve this equation for $m$, we measured $a_A = 5.828$ Å by X-ray diffraction (see supplementary Fig. S3). Furthermore, the change of volume at the martensitic transformation can be as low as 0.06 %[37], making volume conservation a good approximation: $a_A^3 = c_T \cdot a_T^2$. This introduces a dependence of $c_T$ and $a_T$ and allows us to reduce the number of variables further, leading to the $c_T/a_T$-ratio as our key parameter.

In general, the resulting $m$ is not an integer number. In a descriptive picture, this means that a small gap occurs between $(n + m)$ austenitic building blocks on one side of the habit plane and the sequence of $n$ dark blue and $m$ light blue tetragonal building blocks on the other side. For $c_T/a_T = 1.205$ of our particular sample (cf. section: Bonus level), the resulting $m$ has to be rounded up to the next integer. This gives $m = 5$, and therefore the $(5\bar{2})_2$ stacking of 14M is expected and observed.

This ordered arrangement of nano-TBs allows introducing a larger unit cell with a monoclinic symmetry (sketched in white in Fig. 3d). Only half of the cell is shown in the $a_{MM}$ direction due to the limited space. All four lattice parameters of this modulated unit cell can be calculated directly from the $c_T/a_T$-ratio by elementary geometry (see Methods section) using the volume conservation from above. For the particular sample, we obtain: $a_{MM} = 29.60$ Å, $b_{MM} = 4.29$ Å, $c_{MM} = 5.48$ Å and $\gamma_{MM} = 85.4°$. As interaction energy can stabilize a modulated unit cell[25], we consider it as a stable building block for all following levels.

**Level 2: Formation of an $(a/b)_{MM}$-laminate**

On this level, we have to deal with the remaining small length difference between the unit cell of modulated martensite and austenite, described at the end of level 1. With increasing number of unit cells, this difference accumulates and results in an increasing elastic energy. To reduce this energy, the same mechanism as described for the tetragonal blocks is used: the introduction of TBs. This results in an alternating orientation of the long and short axes of the modulated cell connected by $(a/b)_{MM}$-TBs (marked red in Fig. 3d). This type of TB had recently been described experimentally[38] as well as theoretically by the concept of ordering nanotwins[25]. An $(a/b)_{MM}$-TB connects two variants V1 and V2 with interchanged ratios of the light and dark blue orientations, respectively. An appropriate length ratio $\lambda$ of both variants within an $(a/b)_{MM}$-laminate allows to adapt exactly to the habit plane (for the particular sample, $\lambda = 0.875$, calculation described on level 3). As this TB coincides with a nano-TB, they are expected to have only a small excess energy in the same order as the interaction energy. This makes $(a/b)_{MM}$-TBs, out of all theoretically possible TBs[29], the most favourable to form an exact habit plane[25]. In agreement with this, they are also observed for 10M martensite with a very narrow spacing[39] showing even some refinement[40].

**Level 3: Nucleation of martensite**

The aim for this level is to use the $(a/b)_{MM}$-laminate as a building block to construct martensitic nuclei. Nucleation requires the encapsulation of a volume of martensite within the austenite using phase boundaries that have a minimum of interface energy. This aspect cannot be solved completely on level 2, as one $(a/b)_{MM}$-laminate only forms one habit plane. However, as habit planes slightly deviate from $\{1\,0\,1\}_A$ in Ni-Mn-Ga, a volume of martensite can be encapsulated by combining several of them[7]. The simplest solution is the combination of eight habit planes in the shape of a diamond as sketched in Fig. 4a, which was also directly observed by in situ experiments[7]. Each of the eight habit planes connects austenite with a particular orientation of an $(a/b)_{MM}$-laminate. To illustrate this nesting of building blocks, the inset depicts the orientation of nano- and $(a/b)_{MM}$-TBs inside one laminate. All laminates can be transformed into each other by mirroring them along the planes going through the middle of the diamond, which allows identifying these "midribs" as TBs. Their interface energies must be provided to enable nucleation. As they occur at the mesoscale, we call them mesoscopic TBs.

For the precise construction of a nucleus, we apply non-linear continuum mechanics, derived by Ball & James[17]. A comprehensive description can be found in ref.[24]. This continuum theory uses lattice parameters and symmetry of both phases as input parameters and predicts the orientations of all variants forming the $(a/b)_{MM}$-laminates as well as the habit planes. Thus, this theory suits our approach of nested building blocks, as the lattice parameters of the modulated unit cell were already obtained at level 1 using the $c_T/a_T$-ratio. Furthermore, due to the energetic arguments favouring $(a/b)_{MM}$-TBs at the habit plane (cf. level 2), it is sufficient to consider only these laminates. Continuum mechanics confirms that the midribs of a diamond are compatible mesoscopic TBs. According to the common nomenclature[29], they can be identified as two type I TBs (the first one is marked cyan and the second one coincides with the austenite $(1\,\bar{1}\,0)_A$ plane in Fig. 4a) and a modulation TB (yellow). Furthermore, this theory gives the orientation of the habit planes enabling us to deduce the small opening angles $\alpha$ and $\beta$ of a diamond. For a better comparison with the observed microstructure, the diamond is viewed along its longest axis in Fig. 4b. In this viewing direction, the vertical type I TB corresponds to the one that lies within the austenite $(1\,\bar{1}\,0)_A$ plane.

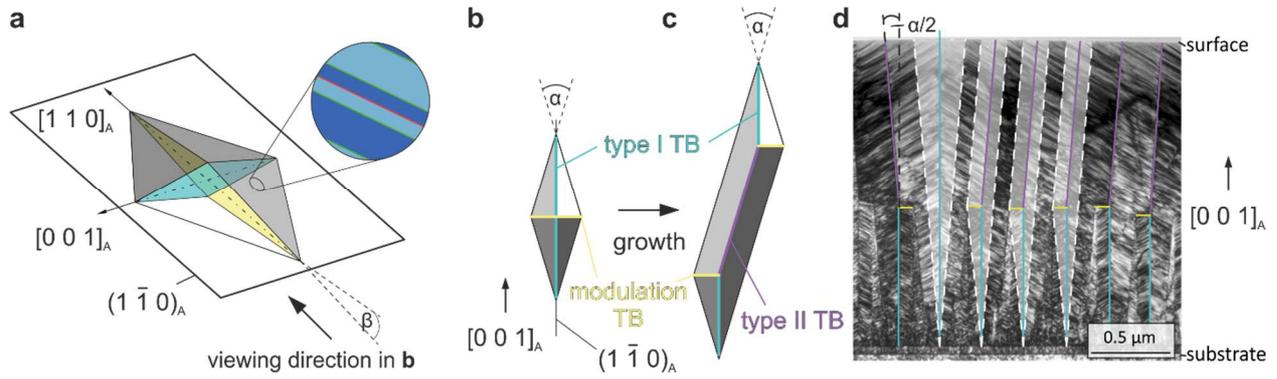

**Fig. 4 On level 3, (*a/b*)$_{MM}$-laminates are used as building blocks for diamond and parallelogram nuclei. a** Sketch of a diamond shaped nucleus, consisting of eight differently aligned (*a/b*)$_{MM}$-laminates. As illustrated by the zoom-in, each of them includes nano-TBs (green lines) and (*a/b*)$_{MM}$-TBs (red line). The midribs of this diamond consist of two type I TBs (the first one is marked cyan, the second one correlates with the austenite $(1\bar{1}0)_A$ plane) and a modulation TB (yellow). **b** Frontal view of the nucleus as indicated in (**a**). The visible type I TB is the one that lies in the austenite $(1\bar{1}0)_A$ plane. **c** The growth of a diamond into a parallelogram nucleus introduces an additional type II TB (purple). **d** Overlay of a section of the TEM cut (from Fig. 2b) with parallelogram nuclei. To reduce the total interface energy, only the bottom part of the parallelogram occurs in a thin film. As this sample has fully transformed, the dashed white lines mark the position of the habit plane between austenite and martensite in an intermediate state during the transition, when the nuclei just reach the substrate. By merging with neighbouring nuclei during growth, identity boundaries form, which are therefore not visible. In all sketches, the angles *α* and *β* are increased in order to improve visibility of all features.

By selectively extending some of the eight habit planes, they can form a more complex, parallelogram shaped nucleus geometry[7] shown in Fig. 4c. This was confirmed recently also for bulk samples[41]. Within such a nucleus, an additional type II TB (purple) forms, which is known for its extraordinarily low twinning stress[42]. The type II TB is slightly inclined by the angle *α*/2 from the $(1\bar{1}0)_A$ plane. In situ experiments revealed the transformation from a diamond into a parallelogram, which is driven by a decrease in total free energy. Thereby, the transformed volume can further increase once a diamond reaches incompatible boundaries during growth[7].

Following the concept of nested building blocks, the geometry of both types of nuclei is fully determined by the $c_T/a_T$-ratio. For the particular sample with a $c_T/a_T$-ratio of 1.205, we obtain *λ* = 0.875, *α* = 10.5° and *β* = 3.1°. This high value of *λ* means that the habit planes are dominated by the variant V1 (cf. Fig. 3d) and only a small fraction of V2 occurs. The characteristic geometries and angles can also be found in the cross-section TEM image (Fig. 4d) suggesting that the shape of these building blocks is conserved during growth. The average angle between the type I and type II TB (*α*/2) was measured to be around 4.5° using the TBs shown in Fig. 4d. The dashed white lines visualize the habit plane in an intermediate state during growth, when the nuclei just reach the substrate. A nucleus can save some of its total mesoscopic TB energy by moving partly out of the film. This agrees with the shown cross-section, where one tip of the nucleus is not visible. All nuclei consist of the same set of eight variants and diagonally opposing variants within one nucleus are identical. Thus, when they meet often no TB is visible after the coalescence of two neighbouring nuclei – the previous habit planes become an "identity boundary".

**Level 4: Growth of martensite towards a herringbone laminate**

The growth of martensite is driven by the lower free volume energy of the martensite phase compared to austenite below the transformation temperature. Accordingly, nuclei grow as large as possible, but this growth is limited by incompatible boundaries, which would require an additional interface energy. Examples of incompatible boundaries are grain boundaries, the interface to the substrate, or regions that had already transformed to the martensite before. When a diamond meets an incompatible boundary during growth, it may transform to a parallelogram for further growth[7], but when this parallelogram meets the next incompatible boundary afterwards, it cannot grow further. This determines the length scale, which is not predicted by continuum mechanics. Thus, many diamonds and parallelograms are required to transform polycrystalline materials or films on rigid substrates compared to a single crystal.

A way to transform most of the volume while avoiding incompatible boundaries is sketched in 2D within the blue shaded area of Fig. 5a. Diamond and parallelogram nuclei are used as building blocks to assemble a self-accommodated herringbone laminate. A diamond nucleus has two equivalent possibilities to transform into parallelograms. Therefore, herringbone TBs (orange) are introduced, which connect both parallelogram orientations. These TBs therefore originate from the spontaneous symmetry reduction when transforming diamonds to parallelograms. As these nuclei originate from the same $\{1\ 1\ 0\}_A$ plane, they fit together because they are surrounded by the same set of habit planes. The microstructure of a herringbone laminate is characterised by the angle $\alpha$ (cf. Fig. 4c) of its building blocks. In order to enable a comparison with experiment, a Fourier transformation of an area containing only type X martensite was done (supplementary Fig. S4) and gives $\alpha = 8.8°$ ($\alpha/2 = 4.4°$). This nearly matches the $\alpha/2 = 4.5°$ measured for the type Y martensite (Fig. 4d) and is in reasonable agreement with the value of $\alpha/2 = 5.25°$ obtained from continuum mechanics for $c_T/a_T = 1.205$.

In thin films, the growth of the nuclei is limited by the substrate. Due to the constant film thickness, a very homogenous spacing of mesoscopic TBs is observed in our single crystal like films. In case of a polycrystalline sample, typically the size of the mesoscopic building blocks becomes smaller when approaching an incompatible grain boundary[43]. Furthermore, the shape of the nuclei appears to be different in each grain. This is because each grain has a distinct orientation and therefore a different cross section of the three dimensional nuclei is visible.

As the tips of the nuclei forming the herringbone laminate have their characteristic angles $\alpha$ and $\beta$, a gap remains when they reach the substrate, as can be seen for the type Y cross-section in Fig. 4d and for the type X martensite in Fig. 5c. This gap can be filled with another nucleus, because the new nucleus has the complementary angle. While in bulk samples these new nuclei can have a similar size, this is not the case for thin films, where the remaining space towards the substrate is smaller. Accordingly, the nuclei towards the substrate become smaller and smaller. During this refinement of the martensitic microstructure, the ratio between volume and TB area of the nuclei decreases, which is similar to classical branching towards the austenite-martensite interface[44]. As mesoscopic TBs require an additional excess energy[45], we expect that a higher undercooling below the transformation temperature is necessary to fill the remaining, small regions. This refinement is also observed at incompatible macroscopic TBs (Fig. 5b).

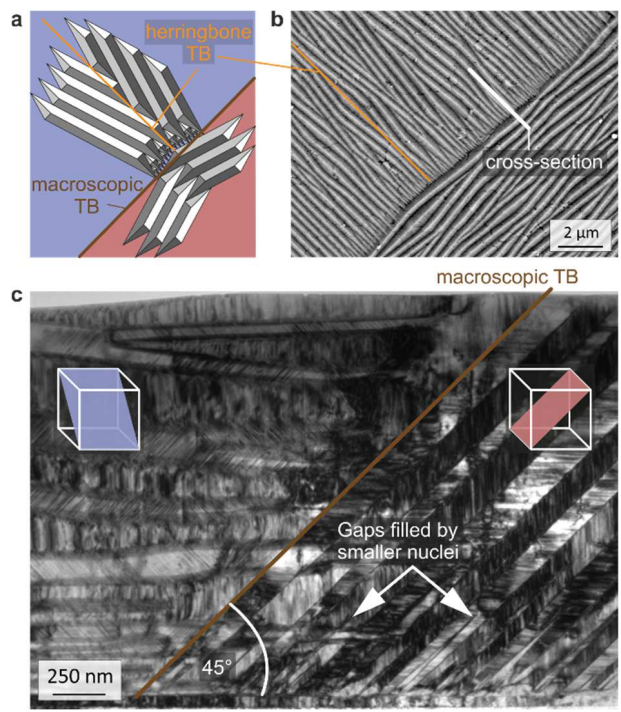

**Fig. 5 On level 4, coalescence of parallelogram shaped nuclei results in a herringbone laminate, which is the building block for level 5. a** Diamonds and parallelograms on the same {1 1 0}$_A$ plane are the building blocks for a self-accommodated herringbone laminate (blue background) incorporating herringbone TBs (orange). When nucleation also occurs on a different {1 1 0}$_A$ plane (red), a macroscopic TB (brown) forms where both herringbone laminates meet. **b** SEM micrograph of the type X martensite structure (area marked in Fig. 2a). **c** TEM cross-section at the position marked in (**b**). The macroscopic TB separating both herringbone laminates is inclined by 45°. The orientation of the associated austenite {1 1 0}$_A$ plane is sketched for the left part in blue and for the right part in red (correlating with **a**). The nuclei are inclined according to the sketched planes. In the cross-section, a small angle remains between the nuclei. These gaps are filled up by smaller nuclei refining the martensitic microstructure towards the substrate. This figure without the overlays is given in the supplementary (Fig. S5).

**Level 5: Macrotwinning**

Here, the herringbone laminate is used as a building block. We follow the same approach as already used at level 1, where the laminate of nano-TBs forms the modulated unit cell. At the nano-scale, we could even calculate symmetry and lattice parameters of this building block. For the macroscale, we give a more general description. The habit planes forming the martensitic nuclei of the herringbone laminate in Fig. 5a (blue part) are all close to one particular {1 0 1}$_A$ plane sketched in blue in Fig. 5c. Due to the cubic symmetry of the austenite, there are six crystallographically equivalent {1 0 1}$_A$ planes. In principal, the herringbone laminate can occur on any of these six planes, leading to six different orientations. As the martensitic phase transition simultaneously starts in different areas of the sample, differently oriented laminates can form. When two of them meet, their growth will stop, and a macroscopic boundary forms between them. This is depicted in Fig. 5a, where the second laminate (red) is rotated by 90° in-plane. These two laminates are connected by the same symmetry operation that also connects the two austenite {1 0 1}$_A$ planes on which they form. Hence, the connecting boundary is quite similar to a classical TB, and we consider it appropriate to call them macroscopic TBs. We propose that the spacing between them originates from the distribution of microstructural defects that can act as heterogeneous nucleation sites for the martensite.

Macroscopic TBs are incompatible because the diamonds and parallelograms forming within each herringbone laminate do not fit together with diamonds forming on a different herringbone laminate. Thus, to reduce the gap at the macroscopic TB, faceting can occur, smaller nuclei are introduced (cf. Fig. 5a,b), or even some residual austenite may remain well below the transition temperature. Such a disturbed region close to a TB, however, is not unique to macroscopic TBs and observed in mesoscopic TBs[46], too.

**Bonus level: Reciprocal space**

Diffraction experiments are decisive when twinning at the nanoscale occurs, creating a modulated structure[30]. Over the years, many studies were performed with an increasing number of fit parameters used to describe the diffraction pattern[22,30,47]. With our scale-bridging approach, one key parameter is sufficient: the tetragonal distortion ($c_T/a_T$) of the smallest building blocks, which is the well-known parameter describing the transformation along the Bain distortion[48]. In addition to this key parameter, the lattice constant of the austenite $a_A$ (see supplementary Fig. S3) is required, assuming volume conservation at the transformation.

We examine the same epitaxial film already used for the real space analysis with its MgO substrate acting as a fixed reference frame. Therefore, we can compare arrangement and orientation of the tetragonal building blocks at the nanoscale. Reciprocal space maps (RSM) are used to verify our model of nested building blocks. This method has the advantage that it probes the sample volume, allowing us to obtain statistically relevant data from an X-ray spot of several mm².

To calculate the RSM, we use the $c_T/a_T$-ratio as key parameter and consider the five levels of building blocks. This is described in the methods part together with further details on the device setup. The comparison between measured and calculated RSM in Fig. 6 reveals a good agreement of diffraction positions and intensity. This demonstrates that our approach of nested building blocks is suitable to describe the reciprocal space of a hierarchically twinned microstructure. For the calculated RSM, an optimized value for $c_T/a_T = 1.205$ is obtained, as shown in the inset of Fig. 6.

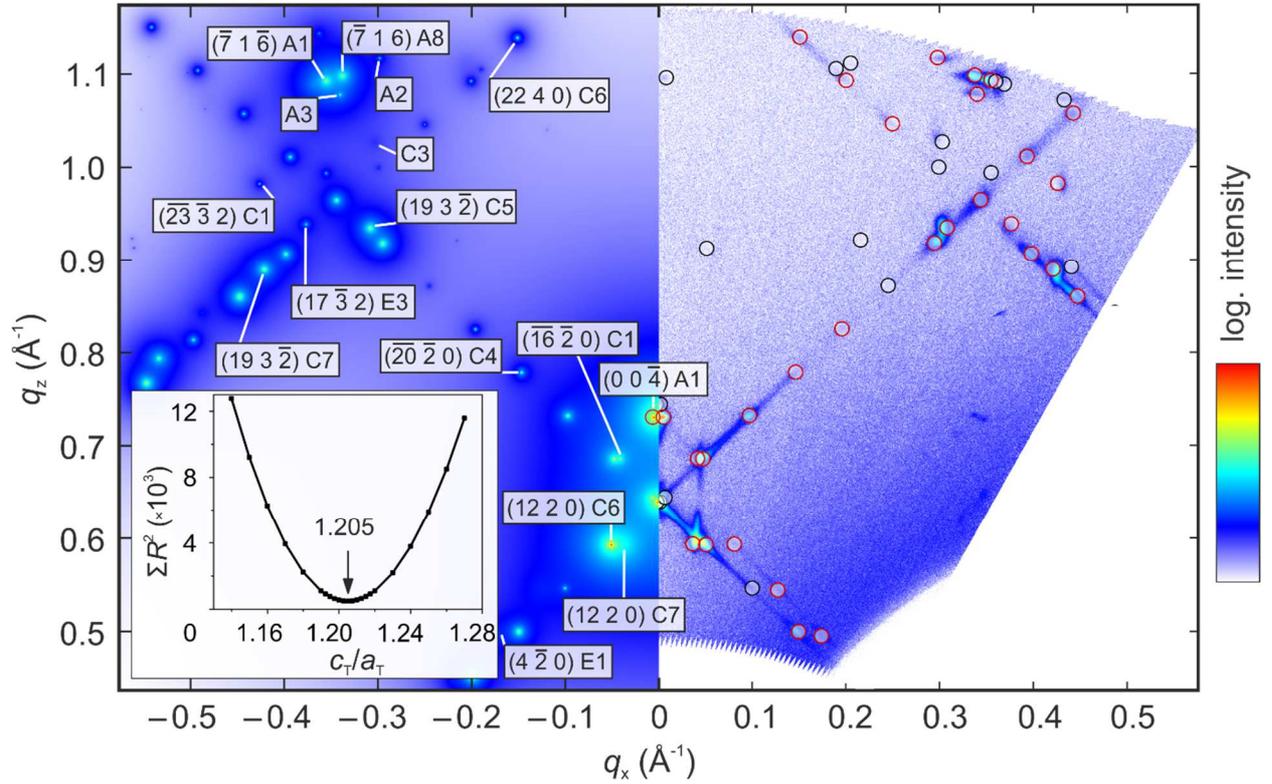

**Fig. 6 Reciprocal space map (RSM) of a hierarchically twinned microstructure.** Measured (right) and calculated (left) RSM along the in-plane ($q_x$) and out-of-plane ($q_z$) directions agree well for an optimized $c_T/a_T$-ratio of 1.205, as summarized in the inset. It depicts the sum of the squared variation $R^2$ between 30 measured and calculated peak positions when varying the $c_T/a_T$-ratio. The calculated peak positions are marked in the measured RSM with black and red circles. The red circles indicate the peaks used for the determination of the $c_T/a_T$ ratio. Only the majority sets A, C and E are included, see Table 1 in the methods section for further details. Indexing of the reflections is described there as well.

**Preview: Microstructure design**

To optimise the functional properties of materials showing martensitic transitions, designing the microstructure is a crucial point. We therefore look at the levels introduced before and describe how they can be designed. At level 0, the composition can be used to vary the number of valence electrons[49], which decides on the formation of the different modulated/non-modulated phases[50]. We can trace this back to a variation of the $c_T/a_T$-ratio of the tetragonal building blocks as introduced previously[16]. At level 1, this ratio decides which modulated phase forms. Depending on the type of phase, different stress and strain values can be obtained, which allows selecting different properties for actuation by magnetically induced reorientation[51]. In addition, the type of phase is also critical for all caloric effects, since a modulation with a narrow spacing of nano-TBs is necessary to minimise hysteresis losses[52]. On level 2, the formation of $(a/b)_{MM}$-TBs is an aspect which is still in the focus of current research[39]. Our investigation proposes that the amount of $(a/b)_{MM}$-twinning can be controlled by the $c_T/a_T$-ratio, which will allow to examine the impact of these TBs systematically. At level 3 and 4, it is decisive for actuation to obtain type II instead of type I TBs as they exhibit a twinning stress lower by one order of magnitude[42]. Since levels 3 and 4 are governed by nucleation and growth respectively, we propose two measures to control the formation of the herringbone laminate. As a permanent measure, we suggest to utilise the recently identified segregation tendency of Heusler alloys[53] for the formation of defects by precipitations. Depending on size and shape, precipitates can either facilitate nucleation or hinder growth by pinning. As a reversible measure, we propose to adjust cooling rate and temperature gradient for the transition, which allows controlling the kinetics of nucleation and growth on level 3 and 4 separately. Moreover, our concept of nested building blocks as a whole paves the way for a microscopic understanding of the different transformation kinetics observed in martensites, which can be both, athermal and isothermal[54]. This will be decisive for all caloric applications, as the underlying cooling cycle should run at frequencies as high as possible[55]. Furthermore, the size of diamonds and parallelograms on level 4 is limited by incompatible boundaries, which means that they can be controlled for example by grain size, or film thickness in thin films[45]. Thereby, it is for example possible to shift the temperature of the martensitic transition[56]. Lastly, to eliminate macroscopic TBs on level 5, external stress or magnetic fields can be used as they favour particular orientations. This is decisive for magnetically induced reorientation, as incompatible macroscopic TBs hinder any actuation[57]. Overall, our approach of nested building blocks on five levels determines most features of the martensitic microstructure based on the $c_T/a_T$-ratio from level 0. Nonetheless, we could identify a few paths, which can be utilised to design the remaining features of the hierarchical microstructure.

**Conclusions: Quest fulfilled**

Our approach of nested building blocks explains the deeply hierarchical microstructure of martensitic materials. We are able to describe this microstructure on all length scales by creating five consecutive levels of hierarchical building blocks. To proceed from one level to the following, a larger building block is created from smaller ones by twinning as summarized in Fig. 1. Each level is required to solve a particular constraint of martensitic transformations: the formation of a phase boundary on level 1, fine adjustment of strain at this phase boundary (level 2), nucleation (level 3), growth of martensite (level 4) and level 5 due to the symmetry of the austenite. Adapting each of these constraints requires a local energy minimisation at every level instead of a global minimisation. Since each level builds upon the previous one, the complete martensitic microstructure can be constructed using the $c_T/a_T$-ratio of the smallest, tetragonal building block as the key parameter for our model system Ni-Mn-Ga. Furthermore, we could identify the few remaining possibilities to design such a martensitic microstructure (cf. section: "Preview"). A transfer from our model system towards other Heusler alloys is straightforward as they have the same structure. However, our approach of nested building blocks can be applied to other martensitic materials as well. Therefore, the different crystal symmetry has to be considered, which defines the initial building block. We expect that our approach is also applicable to other functional materials like ferroelectrics and multiferroics, where similar diffusionless transitions occur. This can help to further design the microstructure of the aforementioned materials considering that a hierarchical microstructure is crucial for the functional properties.

**Methods**

*Experimental methods*

The 2 µm thick epitaxial Ni$_{48}$Mn$_{33}$Ga$_{19}$ film was grown by magnetron sputter deposition on a single crystalline MgO (0 0 1) substrate heated to 300 °C with a 70 nm Cr-buffer layer underneath. The sample surface was imaged by SEM (FEI Helios NanoLab 600i) using backscattered electron contrast. This dual (ion and electron) beam instrument was also used to prepare two FIB cross-section lamellas for TEM investigations. For the characterisations on the mesoscopic length scale, bright-field TEM was conducted using an FEI Tecnai G2 microscope (ThermoFisher Scientific Comp.) operated at 200 kV acceleration voltage. For nanoscale (atomic) imaging, aberration-corrected high-resolution TEM was carried out using a double-corrected FEI Titan$^3$ 80-300 microscope (ThermoFisher Scientific Comp.) operated at 300 kV acceleration voltage.

The TEM images in Fig. 2b,c, Fig. 4d and Fig. 5c were slightly rotated to align the interface between substrate and film parallel to the picture border and then cropped to a rectangle. Brightness and contrast of Fig. 2b,c and Fig. 4d were slightly optimised, uniformly for the whole image.

Reciprocal space maps (RSM) are a well-established technique to analyse epitaxial thin films as they provide the scattered intensity along a planar cut through the reciprocal space. The cut can be defined by the sample orientation and is almost two dimensional, just limited by the device apertures. Measurements were performed on a Philips X'Pert X-ray device with a four circle goniometer using Cu-Kα radiation and an area detector (Malvern Panalytical PIXcel3D). For each measurement, an in-plane rotation angle of the sample ($\varphi=45°$) and a tilting angle perpendicular to the measurement plane ($\psi = 3.5°$) were specified. The used measurement range was between 0.6° and −28.2° for $\omega$ (sample tilt in beam direction) and between 22.5° and 64° for $\theta$ (incident angle).

*Theoretical methods*

According to the textbook of Bhattacharya[24], we used nonlinear elasticity based continuum mechanics. We included the considerations described on the levels 1…5 as follows:

According to level 1, nanotwinning is used to construct the modulated unit cell. As the spacing of this cell is in the order of the wavelength used for diffraction, this results in superstructure reflections. The lattice parameters of the modulated unit cell are calculated from the nano-twinned tetragonal building blocks: The following formulas are a generalization from the adaptive concept (cf. ref.[35]):

$$b_{MM} = \frac{a_A}{2 \cdot \sqrt[3]{c_T/a_T}} \cdot \sqrt{(c_T/a_T)^2 + 1}$$

$$c_{MM} = \frac{a_A}{\sqrt[3]{c_T/a_T}}$$

$$a_{MM}' = \frac{b_{MM}}{2} \cdot \sqrt{m^2 + n^2 - 2mn \cdot \cos(4\arctan((c_T/a_T)^{-1}))}$$

$$\gamma_{MM} = 2 \cdot \arctan((c_T/a_T)^{-1}) + \arccos\left(\frac{a_{MM}'}{m \cdot b_{MM}} + \frac{b_{MM}}{m \cdot a_{MM}'} \cdot \left(\frac{m^2}{4} - \frac{n^2}{4}\right)\right)$$

As the length of $a_{MM}$ depends on the chemical order, one has to use $a_{MM} = a_{MM}$', if $m + n =$ even, otherwise $a_{MM} = 2a_{MM}$'.

At level 2, $(a/b)_{MM}$ twinning occurs. Accordingly, we consider a set of variants V1 and V2 (cf. Fig. 3d) with a volume ratio ($\lambda$) forming the habit plane to the austenite. This ratio is calculated applying continuum mechanics using $a_A = 5.828$ Å as lattice parameter for the austenite (cf. measurement shown in supplementary Fig. S3) while simultaneously solving the twin and habit plane conditions. The number of resulting variants was reduced to 48 by applying the constraint that only $(a/b)_{MM}$-TBs coinciding with nano-TBs form directly at the habit plane, as described on level 2 in the main part of the paper.

Nucleation at level 3 is considered by taking the particular tilt and rotation of the martensitic variants into account, as the orientation of the habit planes can be calculated by continuum mechanics, too. On level 4, only the size of the laminates forming the habit planes changes during growth. Therefore, no further adjustments have to be done. Finally, we consider level 5 as all possible variants from all austenite $\{1\,0\,1\}_A$ planes are included.

For the calculation of the RSM cuts, we re-sorted the $(a/b)_{MM}$ variants, as global diffraction does not probe their local position. They can be divided into six sets characterized by a common out-of-plane direction. As the volume ratio of the variants ($\lambda$) does not equal 1/2, it allows distinguishing between the majority variants and the minority variants. Table 1 summarizes this classification. An exemplary sketch of such a variant set in real space is given in the supplementary (Fig. S6). A more detailed look into how the variant sets are sorted is given in ref.[58]. Out of the six sets, four belong to the type X martensite and two belong to the type Y martensite. This was confirmed by comparing the calculated and the measured RSMs. The ratio between type X and Y varies from film to film. To understand this behaviour, one has to keep in mind that in contrast to bulk materials, not all $\{1\,0\,1\}_A$ planes are equivalent for thin films. Commonly, the direction perpendicular to the substrate differs from the in-plane directions. Accordingly, in epitaxial $(0\,0\,1)$ films, diamonds forming on the perpendicular $\{1\,1\,0\}_A$ planes (type Y) see a different set of boundary conditions compared to the ones forming on the planes inclined by 45° (type X, cf. Fig. 5c). As both types are cut by a $(0\,0\,1)_A$ plane at the surface, they appear different in Fig. 2a, though they consist of the same mesoscopic building blocks: diamonds and parallelograms. It had been suggested that the ratio between type X and type Y depends on biaxial stress, often present in thin films[59]. As variant selection by stress is not included, the present model cannot predict the deviation from the ratio of 2/4 expected from bulk symmetry (cf. supplementary Fig. S1).

**Table 1 Classification of martensite variants grouped into six different sets (A-F) each consisting of eight variants sharing a common out-of-plane direction.**

| Set | Out-of-plane axis (approx.) | Forms $(a/b)_{MM}$ TB | Variant category |
|---|---|---|---|
| **A – type X** | $c_{MM}$ | with B | majority |
| **B – type X** | $c_{MM}$ | with A | minority |
| **C – type X** | $a_{MM}+7b_{MM}$ | with D | majority |
| **D – type X** | $a_{MM}-7b_{MM}$ | with C | minority |
| **E – type Y** | $a_{MM}-7b_{MM}$ | with F | majority |
| **F – type Y** | $a_{MM}+7b_{MM}$ | with E | minority |

For the intensity calculation, the structure factor of every $(h\,k\,l)$ combination in the investigated part of the reciprocal space is computed for all 48 variants. It is assumed that the intensity decays by a Pseudo-Voigt function in all directions. An additional diffraction contribution, possibly originating from a narrow spacing of V1 and V2 inside the $(a/b)_{MM}$ laminate is neglected as well as the interference on all bigger length scales.

The intensity in the two-dimensional cut through the reciprocal space is summed up reflecting the measurement conditions (basic orientation and tilt of the sample). The used laboratory X-ray measurement system is only weakly cropping in $\psi$-direction (perpendicular to the measured plane). This causes diffraction intensity to be collected from an around one degree $\psi$-interval in both directions. To accommodate for that, there were cuts made every 0.025° through the calculated intensity distribution, which were summed up using a weighting in the form of a triangular function. Accordingly, a cut at the $\psi$ angle of the measurement gets the weight one, which decreases until it reaches zero at $\Delta\psi = 1$. The calculated cuts do not account for device parameters and measurement specifics like different intensities in positive and negative $\omega$-direction (corresponding to $\pm q_x$ in the RSM). Thus, the calculations provide a precise peak position but only an estimate on intensity. The simulated images are log-scaled to allow for an easy comparison with the measured intensities. The peaks in Fig. 6 are indexed as follows: The most intense peaks are marked by the ($h\,k\,l$) planes they originate from, followed by a letter corresponding to the set of martensitic variants (cf. Table 1). The terminal number specifies the variant figure in the set ranging from 1 to 8 (cf. Fig. S6 for more details). Due to a lack of space, three peaks are given only by the variant set and number, the full index for these three is as follows: ($\bar{9}\,\bar{1}\,\bar{6}$) A2, ($5\,1\,6$) A3 and ($\bar{23}\,\bar{3}\,2$) C3. In the modulation direction of a single variant, there is at least one indexed peak. The remaining ones can be deduced by increasing or decreasing the $h$ in steps of two, e.g., the upper left peak close to ($\bar{20}\,\bar{2}\,0$) C4 is ($\bar{22}\,\bar{2}\,0$) C4.

To refine our reciprocal space calculations, two assumptions could be changed: First, we considered only the perfect $(5\bar{2})_2$ stacking of a modulated unit cell. However, faults in the stacking sequence can easily occur, as the starting point for the modulated structure is a disordered arrangement just after the phase transition (cf. Fig. 3c,d). In agreement with the measurement shown in Fig. 6, stacking faults will broaden the peaks along the modulation direction and may result in deviations from their equal spacing. Secondly, we assumed a unit cell with complete chemical order. Accordingly, some deviations are expected for the investigated Heusler alloy. However, as the goal of this paper was to use a minimum of parameters for the description of a hierarchical martensitic microstructure, these refinements were not done here.

**Availability of computer code and algorithm, data and material**

Raw data of the micrographs and sample material as well as the RSM data are available on reasonable request.

**Acknowledgment**

This work was funded by the German Research Foundation (DFG) by project FA453/11 via the Priority Program SPP 1599. The authors acknowledge C. Behler for preliminary TEM micrographs on a similar sample as well as U. K. Rößler, K. Lünser, A. Diestel, S. Engelhardt and A. Waske for helpful discussion.

**Author contributions**

S.S. developed the scale-bridging model to calculate the RSM measurements and wrote the first version of the manuscript. R.N. developed the combination of nano-twinning and continuum mechanics. A.B. deposited the film. D.W. gathered the HR-TEM images. C.D. made the TEM overview images. T.W. prepared the TEM lamellas and took the SEM images. H.S. contributed to the implementation and consequences of the nonlinear elasticity based continuum mechanics. O.H. contributed on sorting the different types of TBs. K.N. supervised the thesis of S.S. S.F. revealed the hierarchical microstructure and wrote the second version of the manuscript. All the authors contributed to the final version.

**Competing interests**

We declare no competing interests.

Supplementary for:
# Building hierarchical martensite

Stefan Schwabe, Robert Niemann, Anja Backen, Daniel Wolf, Christine Damm, Tina Walter, Hanus Seiner, Oleg Heczko, Kornelius Nielsch, and Sebastian Fähler

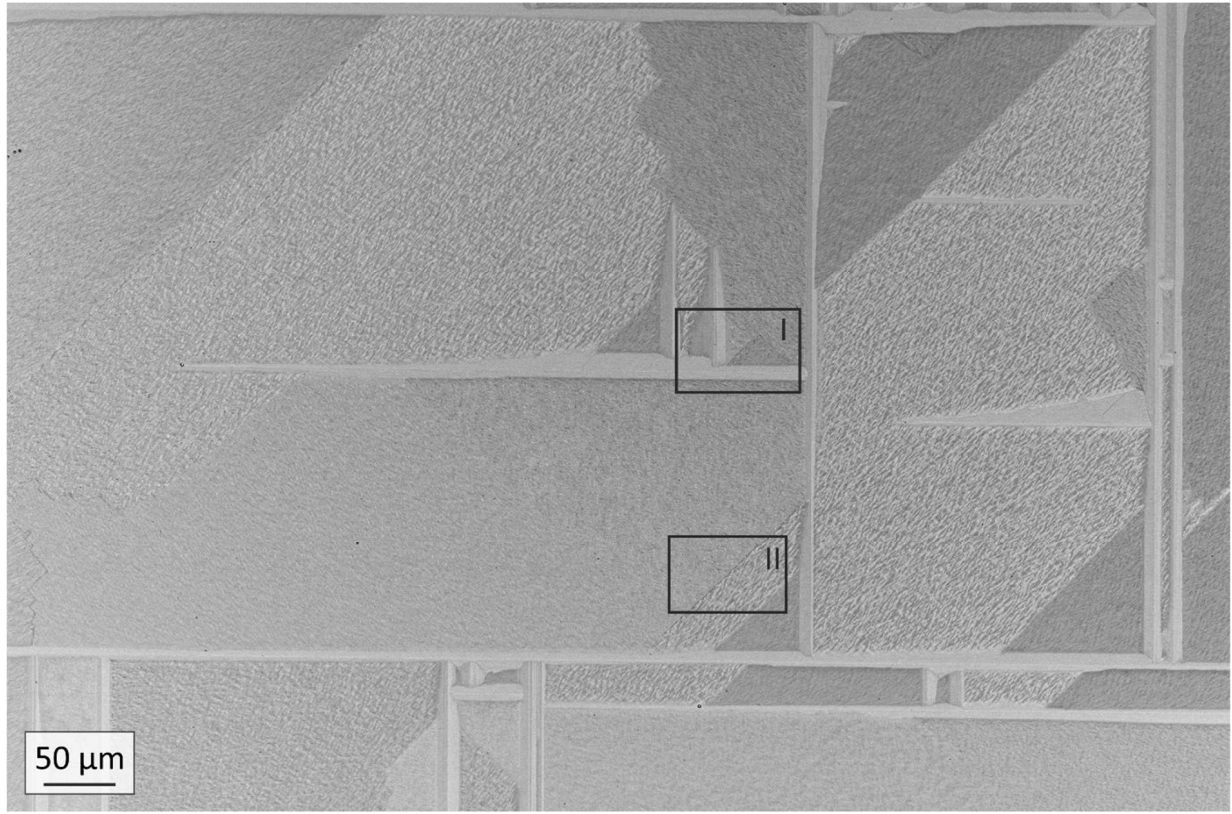

**Fig. S1 Overview illustrating the ratio between type X and type Y martensite.** Top view (SEM with backscattered electron contrast) of a representative sample area. The surface fraction of type Y martensite was determined to be around 9.8 %. The evaluation of a second picture showing a different sample area (same size, not shown) resulted in a fraction of 10.8 % type Y martensite. Therefore on average, we expect the sample to consist of around 10 % type Y martensite with an absolute error of ±2 %. The frame labelled "I" corresponds to Fig. 2a of the main paper. Frame II shows the area discussed in Fig. S4 of the supplementary.

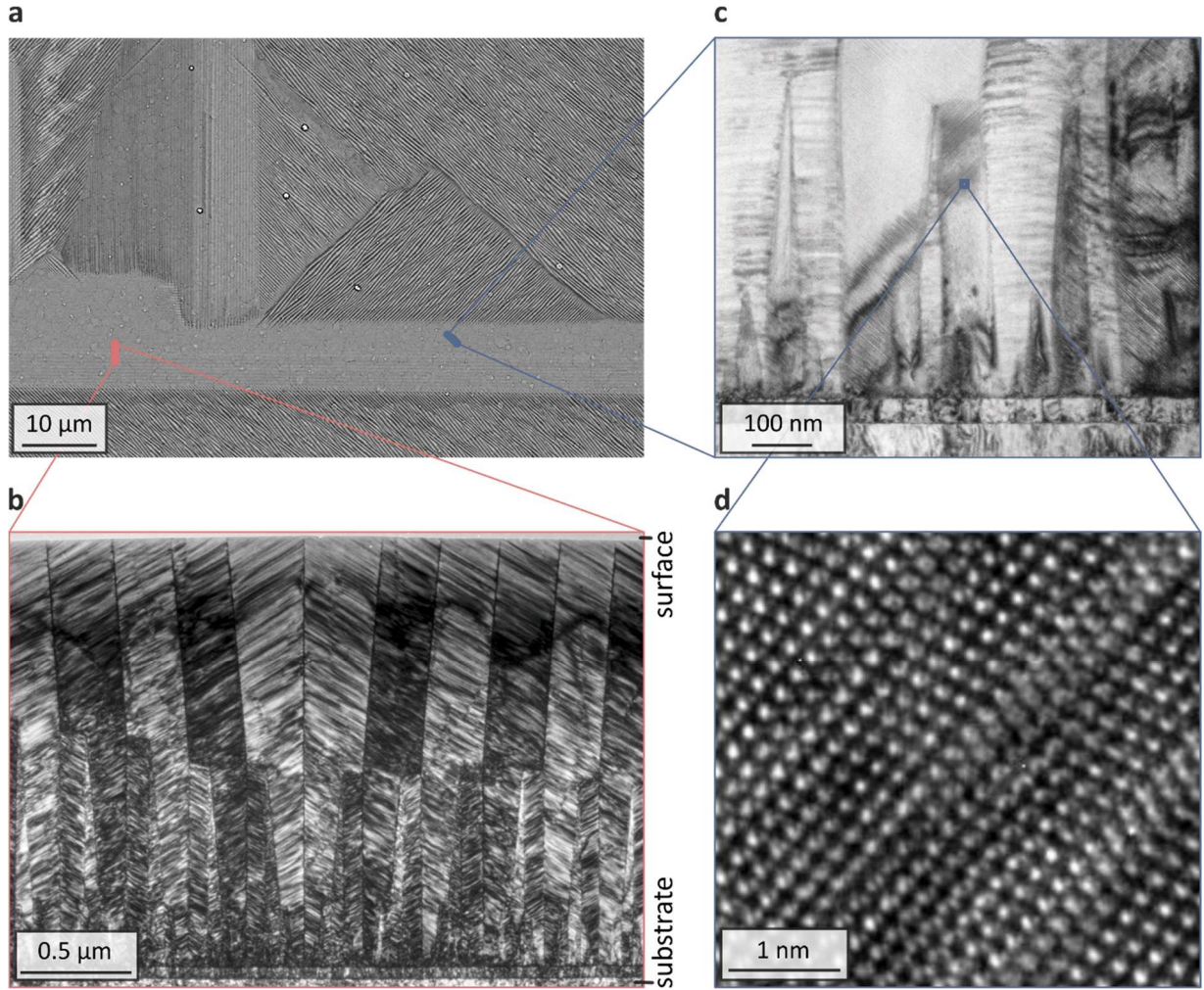

**Fig. S2 Zoom into the hierarchical microstructure without overlays (corresponds to Fig. 2, main paper). a** Macroscopic level observed by a top view SEM (backscattered electron contrast). **b** Mesoscopic scale shown in a TEM cross-section (position marked red in (**a**). **c** A second TEM cross-section cut rotated by 45° with regard to (**b**). **d** HR-TEM image zooming into the area marked with a blue square in (**c**). The figure edges in (**a**) are parallel to $[110]_A$ and $[\bar{1}10]_A$, respectively.

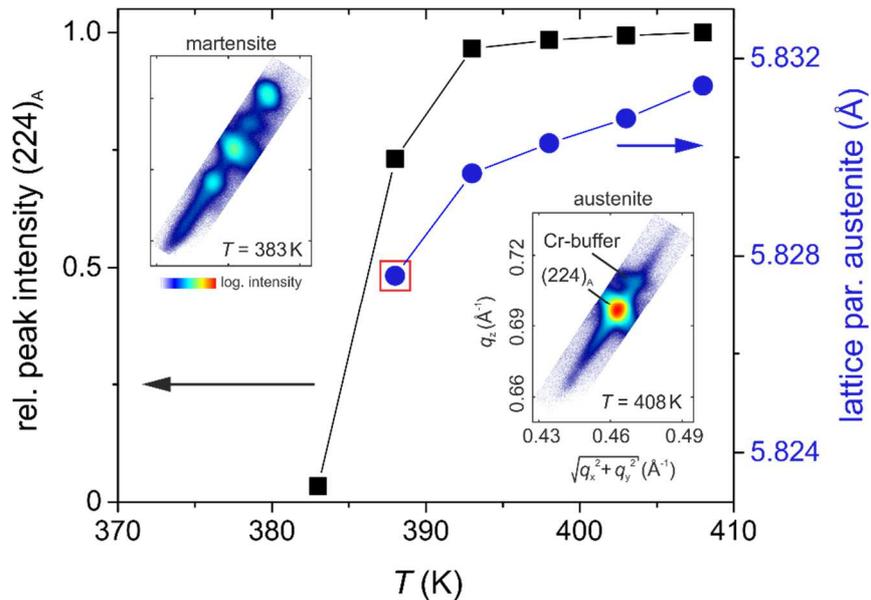

**Fig. S3 Temperature dependent lattice parameter of the austenite ($a_A$) and relative intensity of the (224)$_A$ peak.** $a_A$ was determined by temperature dependent reciprocal space maps (RSM) of the (224)$_A$ and (004)$_A$ peaks. The measurements were done using the device described in the methods section (main paper). Two exemplary RSM are shown as insets for 408 K (austenite) and 383 K (martensite). The axes for the martensitic RSM are conform to the ones for the austenitic RSM. Close to the transition (lowest temperature still showing austenite, marked red), the determined lattice parameter is $a_A = 5.828$ Å.

ii

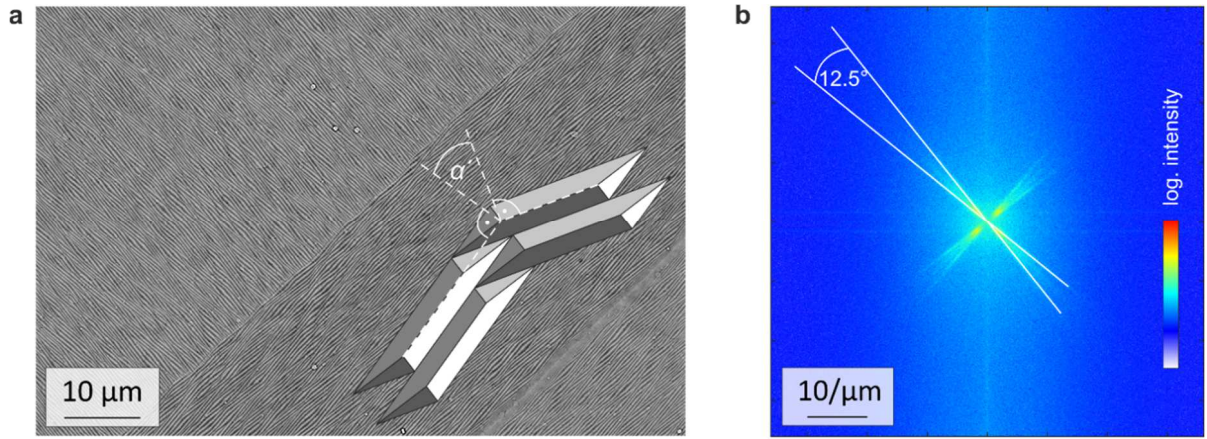

**Fig. S4 Determining the characteristic angle *α'* from a Fourier transformation of the herringbone laminate shown in Fig. S1.** **a** The frame used for the Fourier transformed image depicted in **b**. The schematic parallelogram nuclei shown in the overlay in (**a**) illustrate the orientations occurring in this area as well as the angle *α'*. As type X martensite is inclined by 45° with respect to the surface (cf. Fig. S5), the obtained value *α'* = 12.5° was corrected, resulting in *α* = 8.8°. This allows a comparison with the continuum mechanics calculations. The average periodicity of the TBs is around 400 nm.

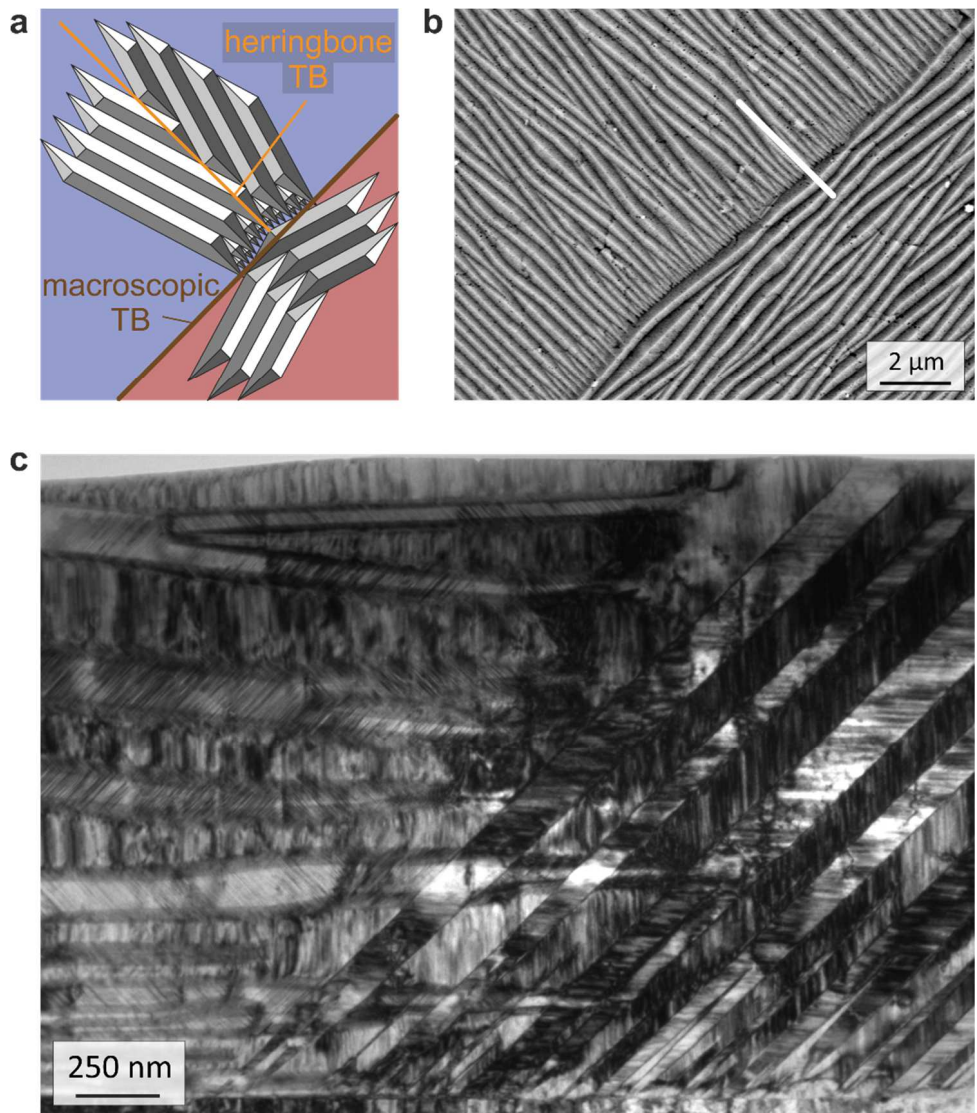

**Fig. S5 Herringbone laminate and type X cross-section without overlays (corresponds to Fig. 5, main paper) a** Diamonds and parallelograms on the same $\{1\,1\,0\}_A$ plane are the building blocks for a compatible herringbone laminate (blue background), which incorporates herringbone TBs (orange). When nucleation also occurs on a different $\{1\,1\,0\}_A$ plane (red), a macroscopic TB (brown) forms where both herringbone laminates meet. **b** SEM micrograph of the type X martensite structure. **c** TEM cross-section at the position marked white in (**b**).

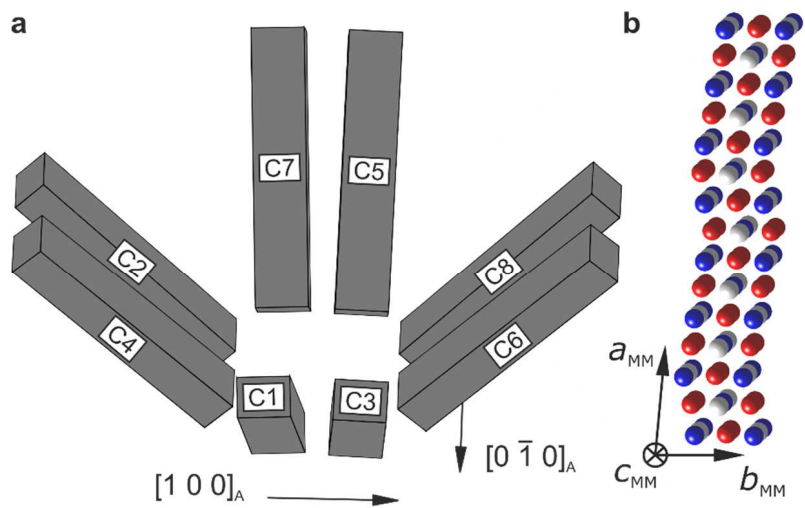

**Fig. S6 Exemplary sketch of eight symmetry-equivalent variants. a** Variants belonging to set C as described in Table 1. The arrangement corresponds to an austenite cell, which is aligned according to the given axes. Each grey parallelepiped represents a 14M martensite cell as shown in **b.** The red atoms correspond to Ni, the blue ones to Mn and the white ones to Ga. To obtain the orientation of the other sets (A-F), the variants have to be tilted according to the out-of-plane axis assigned in Table 1.